\documentclass[preprint,aps,superscriptaddress]{revtex4}
\usepackage{graphicx}
\usepackage{dcolumn}
\usepackage{bm}
\usepackage{times}

\begin{document}
\title{Weak-wave advancement in nearly collinear four-wave mixing}

\author{C. F. McCormick}
\affiliation{Department of Physics, University of California, Berkeley, CA 94720-7300.}
\author{R. Y. Chiao}
\affiliation{Department of Physics, University of California, Berkeley, CA 94720-7300.}
\author{J. M. Hickmann}
\email{hickmann@loqnl.ufal.br}
\affiliation{Department of Physics, University of California, Berkeley, CA 94720-7300.}
\affiliation{Departamento de F\'{\i}sica, Universidade Federal de Alagoas, CidadeUniversit%
\'{a}ria, 57072-970, Macei\'{o}, AL, Brazil.}

\begin{abstract}
We identify a new four-wave mixing process in which two nearly collinear pump beams produce phase-dependent gain into a weak bisector signal beam in a self-defocusing Kerr medium.  Phase matching is achieved by weak-wave advancement caused by cross-phase modulation between the pump and signal beams.  We relate this process to the inverse of spatial modulational instability and suggest a time-domain analog.
\end{abstract}

\maketitle

Modulational instability (MI) is a common and important nonlinear effect in which a continuous wave breaks up into periodic, localized wave packets.  Its long-term dynamics can lead to Fermi-Pasta-Ulam (FPU) recurrence, in which a nonlinear system returns to its initial energy configuration rather than thermalizing\cite{Fermi1965}.  Interest in MI has increased since the recent experimental observation of FPU recurrence of nonlinear optical pulses in optical fibers\cite{VanSimaeys2001, VanSimaeys2002, Akhmediev2001}.  MI in the spatial domain is known as beam filamentation and is a four-wave mixing process in which the phase matching is provided by cross-phase modulation\cite{Agrawal1990,hickmann1992,Boyd1999}. 

Recent experimental work has demonstrated the existence of a connected spatial effect, photon-photon scattering in the presence of a Kerr nonlinearity\cite{Mitchell2000}. In this experiment, two counter-propagating beams ``collide'' in a rubidium vapor cell and photons scattered at $\pm 90^{\circ}$ from the beam axis are observed to be correlated in time, while photons at other angles are not. These results suggest a momentum-conserving photon-photon scattering process mediated by the Kerr nonlinearity. 

In this paper we discuss a related process, in which the two ``colliding'' beams are in a frequency-degenerate, nearly-collinear configuration. Normally this configuration generates high-order diffracted beams, used for example in dephasing time measurements. We predict that a different kind of four-wave mixing process should be possible with this geometry, in which two photons of the noncollinear pump beams are absorbed and two collinear signal photons are generated in the direction of a bisector line between the pump beams. We demonstrate that through a phenomenon similar to \emph{weak-wave retardation}~\cite{Chiao1966}, phase matching is possible even if all four waves have the same frequency.  We conclude by identifying this process as the inverse of the process that generates standard beam filamentation.

\begin{figure}[tbp]
\centerline{\includegraphics{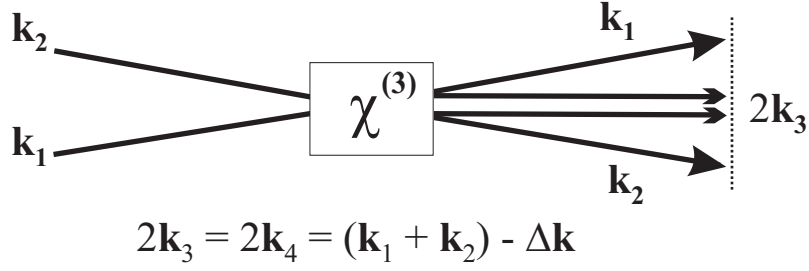}}
\caption{Wave vector configuration for weak-wave advancement.}
\label{configuration}
\end{figure}

We begin by calculating the nonlinear polarization component in the
direction of the signal beam, $P_{4}$. For simplicity we assume that this
process will be stimulated, introducing a small seed field $E_{3}$ in the
direction of the bisector line (see Fig. \ref{configuration}).  The general expression for the polarization of the signal field is

\begin{eqnarray}
P_{4} &=& 3\chi_{xxxx}^{(3)}
\{
|E_{4}|^{2}E_{4} + 2(|E_{1}|^{2} + |E_{2}|^{2} + |E_{3}|^{2})E_{4} \nonumber \\
&& + E_{1}E_{2}E_{4}^{\ast } \exp (i(\vec{k}_{1}+\vec{k}_{2}-\vec{k}_{3}-\vec{k}_{4}) \cdot \vec{z} - i(\omega_{1}+\omega_{2}-\omega_{3}-\omega_{4})t) 
\}
\label{polarization}
\end{eqnarray}

\noindent where $\chi _{xxxx}^{(3)}$ is a component of the third-order nonlinear susceptibility tensor and $E_{j}$ are the (complex) fields associated with the pump ($j=1,2$) and signal ($j=3,4$) beams. The first term in this equation represents self-phase modulation (SPM), the second cross-phase modulation (XPM) and the third four-wave mixing. For a weak signal beam, SPM is negligible. The XPM terms from the pump beams are large and play an important part in the four-wave mixing process, while the XPM term from the signal beam may be neglected. The exponential four-wave mixing term indicates that in order for this polarization to efficiently transfer energy into the signal field $E_{4}$, there must be energy conservation and phase matching between the different wave vectors associated with each field.

In the degenerate case of equal frequencies, this phase matching appears
difficult to achieve. However, the signal wave experiences a decrease in the refractive index generated by the XPM term associated with the pump beams, assuming a negative nonlinear Kerr coefficient (the self-defocusing case). This difference in the refractive index is given by

\begin{equation}
\Delta n=-\frac{3}{4n_{0}}|Re[\chi_{xxxx}^{(3)}]|(|E_{1}|^{2}+|E_{2}|^{2}) 
\end{equation}

\noindent where $n_{0}$ is the linear index of refraction. Consequently, there exists a shortening of the weak-wave momentum vector by $\Delta k=-|\Delta n|\omega_{0}/c$, an effect which we call \emph{weak-wave advancement} in analogy with ``weak-wave retardation'' in the case of the self-focusing sign of the Kerr nonlinearity\cite{Chiao1966}. For a certain parameter range, weak-wave advancement allows phase matching between the four wave vectors. In this range, this four-wave mixing process should transfer energy to the weak signal beam as efficiently as its counterpart.

We now take the special case of degenerate signal fields $E_{3}$ and $E_{4}$.  In this case, we use Eq. \ref{polarization} to calculate the evolution equation for $E_{4}$ in the paraxial and plane wave limits, finding

\begin{equation}
\frac{\partial E_{4}}{\partial z} = i4\gamma P (E_{4} + 0.5 E^{*}_{4}
e^{i(6\gamma P - \Delta k)z}e^{i(\phi_{1} + \phi_{2})})
\label{evolution}
\end{equation}

\noindent where $\gamma \equiv 6\pi |\vec{k}_1|\chi_{xxxx}^{(3)}/n_{0}^{2}$, $P$ is the pump beam power, $\Delta k \equiv |\vec{k}_{1} + \vec{k}_{2} - 2\vec{k}_{4}|$ is the momentum mismatch and $\phi_{1}, \phi_{2}$ are the phases of the pump beams. Taking the self-defocusing case ($\gamma P < 0$) we guess a solution to Eq. \ref{evolution} given by

\begin{equation}
E_{4} = A e^{gz} e^{i(3\gamma P - \Delta k/2)z} e^{i \phi_{4}}
\label{trialsolution}
\end{equation}

\noindent where $A$ is the signal field amplitude, $g$ is a purely real exponential gain coefficient and $\phi_{4}$ is the signal beam phase (for a similar treatment, see reference \cite{Agrawal2001}, p. 392).  Since the phase $\phi_{4}$ is a free parameter, Eq. \ref{trialsolution} is a solution of Eq. \ref{evolution} when the momentum mismatch is in the range  

\begin{equation}
0 \leq \Delta k \leq 6 |\gamma P|
\label{range}
\end{equation}

\noindent with an exponential gain $g$ given by

\begin{equation}
g = \sqrt{3(\gamma P)^{2} - (\gamma P)\Delta k - 0.25(\Delta k)^{2}} \  .
\end{equation}

\begin{figure}
\centerline{\includegraphics{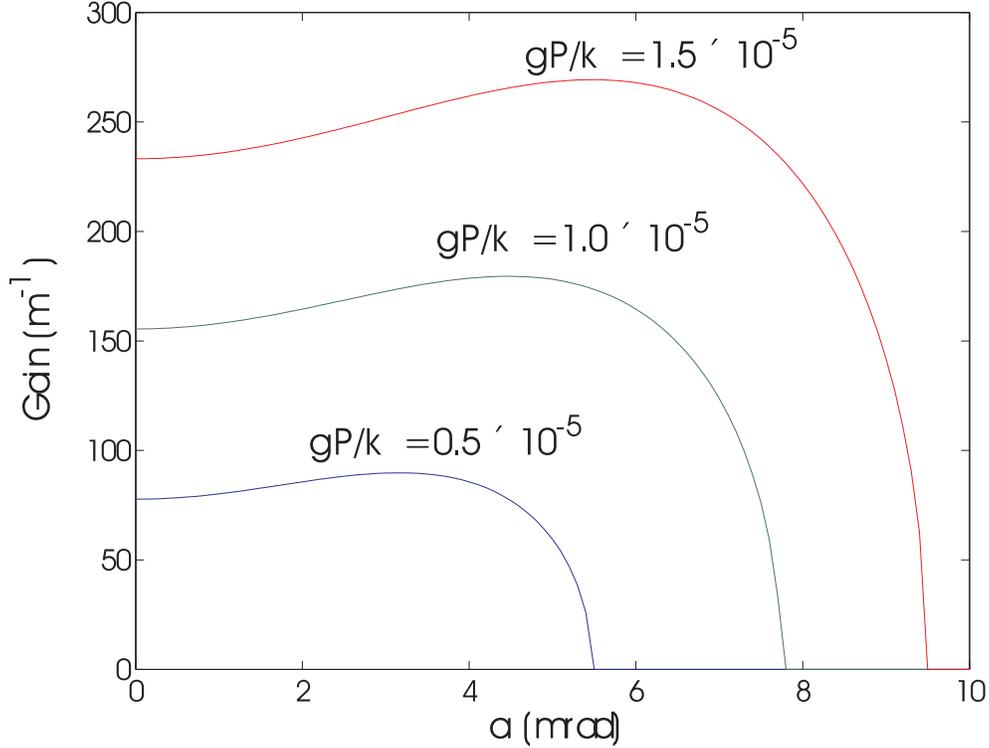}}
\caption{Signal beam gain as a function of collision half-angle $\alpha$ for $\lambda = 700$nm.  Note that $\Delta k = 2|\vec{k}_{1}| (1 - \cos \alpha)$.}
\label{gain}
\end{figure}

The signal beam gain is plotted against the collision half-angle in Fig. \ref{gain} for several values of $\gamma P/k$.  The maximum gain $g = 2|\gamma P|$ occurs when $\Delta k = 2|\gamma P|$.

The geometry considered here has an important difference from the geometry of spatial modulational instability.  Unlike the case of beam filamentation\cite{Chiao1966B,Campillo1973}, the relative phase of the signal and pump beams is under experimental control.  For the optimal choice of $\Delta k$ as above, the signal beam gain is maximized for the phase relation 

\begin{equation}
\Delta \phi \equiv \phi_{1} + \phi_{2} - 2 \phi_{4} = \pi/2
\label{phase}
\end{equation}

\begin{figure}
\centerline{\includegraphics{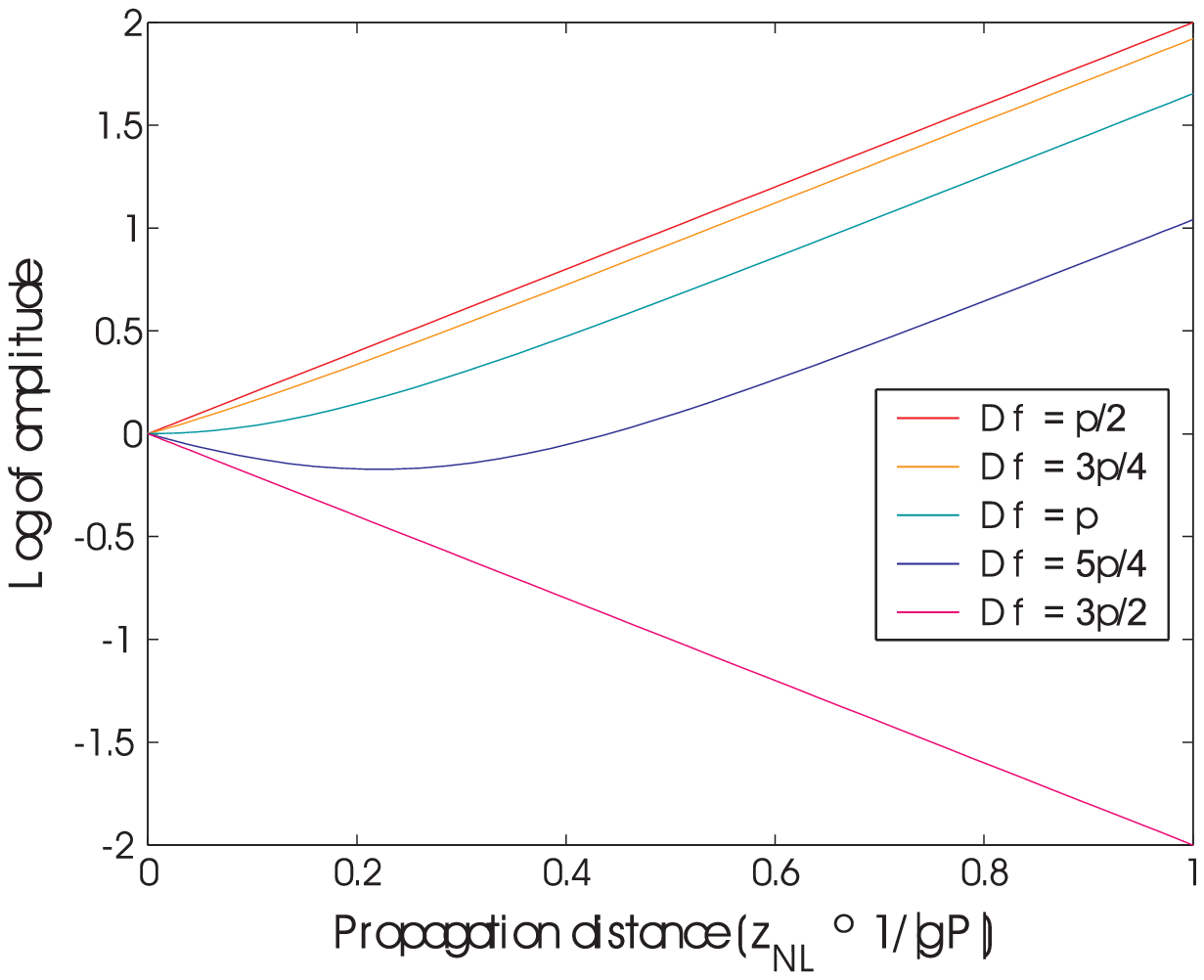}}
\caption{Logarithm of signal beam amplitude as a function of propagation distance.}
\label{gainvphase}
\end{figure}

\noindent For other relative phases the signal beam initially experiences either gain ($0 < \Delta \phi < \pi$) or loss ($\pi < \Delta \phi < 2\pi$). However, we find from numerical solutions to Eq. \ref{evolution} that as the signal beam propagates it accumulates phase so as to satisfy Eq. \ref{phase}.  The signal beam thus eventually experiences gain even if it was initially lossy.  The distance over which this rephasing occurs increases for phases far from $\Delta \phi = \pi/2$, becoming infinite for $\Delta \phi = 3\pi/2$.  It should be noted that the exact phase relation $\Delta \phi = 3\pi/2$ is unstable, and any phase noise will cause $\phi_{4}$ to evolve towards satisfying Eq. \ref{phase}.  Numerical solutions for the log of the signal beam amplitude are plotted against distance in Fig. \ref{gainvphase} for several pump-signal phase relations.

We believe that this gain should be experimentally observable with current materials and laser powers.  We anticipate that the principal difficulty will lie in distinguishing the signal beam from the pump beams, since their only distinguishing feature is the propagation direction.  Given a minimum angular resolution between the beams of $\alpha \approx 1^{\circ}$, satisfying the relation in Eq. \ref{range} implies

\begin{equation}
|\gamma P| / k \geq \frac{\Delta k}{6k} = \frac{1}{3} (1 - \cos \alpha) \approx 5 \times 10^{-5}.
\label{nonlinearity}
\end{equation}

\noindent For a material with $|\chi^{(3)}_{xxxx}| = 10^{-6} \mathrm{cm}^{2}/ \mathrm{W}$, Eq. \ref{nonlinearity} is satisfied for $P \geq 3 \mathrm{W/cm}^{2}$.  Nonlinearities of this order are available in current materials including atomic vapors\cite{Wang2002}.

We recall that weak-wave retardation in degenerate four-wave mixing is
responsible for transverse modulational instability and resulting laser beam filamentation in the spatial domain\cite{Boyd1999}. This behavior agrees with the idea that time-domain modulational instability may be thought of as a four-wave mixing process where the fiber nonlinearity provides phase matching between the different fields\cite{Agrawal2001}. In modulational instability in the time domain, a CW beam breaks up into a periodic pulse train with the simultaneous appearance of associated spectral side bands\cite{Tai1986}. We conjecture that there should exist a time-domain version of the weak-wave advancement Kerr-mediated four-wave mixing effect described here. This effect should occur if we propagate light at two different frequencies $\omega _{0}+\Omega $ and $\omega_{0}-\Omega $ through an optical fiber in the normal dispersion regime.  These two optical frequency components correspond to the noncollinear pump beams in the spatial case discussed here. As a result, a stationary signal at the frequency $\omega _{0}$ should be generated, corresponding to the bisecting signal beam in the spatial case.

In conclusion, we have identified a new four-wave mixing process mediated by the Kerr nonlinearity, and related this process to a different, inverse kind of modulational instability. We believe that this nonlinear scattering process could be at the heart of an experimental realization of a photonic Bose-Einstein condensate, which would require that photons with high transverse momentum (``hot'' photons) be scattered into low transverse momentum modes (become ``cool'' photons)\cite{Chiao2000}.

We thank M. Trassinelli and J. Garrison for helpful discussions. This work was supported in part by the ONR. JMH thanks the support from Instituto do Mil\^{e}nio de Informa\c{c}\~{a}o Qu\^{a}ntica, CAPES, CNPq, FAPEAL, PRONEX-NEON, ANP-CTPETRO.


\begin{thebibliography}{99}

\bibitem{Fermi1965} E. Fermi, J. Pasta, and H. C. Ulam, in \emph{Collected Papers of Enrico Fermi}, edited by E. Segr\`{e} (The University of Chicago, Chicago, 1965), vol. 2 pp. 977-988.

\bibitem{VanSimaeys2001} G. Van Simaeys, Ph. Emplit, and M. Haelterman, ``Experimental demonstration of the Fermi-Pasta-Ulam recurrence in a modulationally unstable optical wave,'' Phys. Rev. Lett. \textbf{87}, 033902 (2001).

\bibitem{VanSimaeys2002} G. Van Simaeys, Ph. Emplit, and M. Haelterman, ``Experimental study of the reversible behavior of modulational instability in optical fibers,'' J. Opt. Soc. Am. B \textbf{19}, 477 (2002).

\bibitem{Akhmediev2001} N. N. Akhmediev, ``Nonlinear physics - Deja vu in optics,'' Nature \textbf{413}, 267 (2001).

\bibitem{Agrawal1990} G. P. Agrawal, ``Transverse modulation instability of copropagating optical beams in nonlinear Kerr Media,'' J. Opt. Soc. Am. B \textbf{7}, 1072 (1990).

\bibitem{hickmann1992} J. M. Hickmann, A. S. L. Gomes, and C. B. de Ara\'{u}jo, ``Observation of spatial cross-phase modulation effects in a self-defocusing nonlinear medium,'' Phys. Rev. Lett. \textbf{68}, 3547 (1992).

\bibitem{Boyd1999} R. W. Boyd, and G. S. Agarwal, ``Preventing laser beam filamentation through use of the squeezed vacuum,'' Phys. Rev. A \textbf{59}, R2587 (1999).

\bibitem{Mitchell2000} M. W. Mitchell, C. J. Hancox, and R. Y. Chiao, ``Dynamics of atom-mediated photon-photon scattering,'' Phys. Rev. A \textbf{62}, 043819 (2000).

\bibitem{Chiao1966} R. Y. Chiao, P. L. Kelley, and E. Garmire, ``Stimulated four-photon interaction and its influence on stimulated Rayleigh-wing scattering,'' Phys. Rev. Lett. \textbf{17}, 1158 (1966).

\bibitem{Agrawal2001} G. P. Agrawal, \emph{Nonlinear Fiber Optics}, 3$^{rd}$ed. (Academic, San Diego, 2001).

\bibitem{Chiao1966B} R. Y. Chiao, M. A. Johnson, S. Krinsky, H. A. Smith, C. H. Townes, and E. Garmire, ``A new class of trapped light filaments,'' IEEE J. of Quant. Elec. \textbf{QE-2} 467 (1966).

\bibitem{Campillo1973} A. J. Campillo, S. L Shapiro, and B. R. Suydam, ``Periodic breakup of optical beams due to self-focusing,''
Appl. Phys. Lett. \textbf{23}, 628 (1973).

\bibitem{Wang2002} H. Wang, D. Goorskey, and M. Xiao, ``Dependence of enhanced Kerr nonlinearity on coupling power in a three-level atomic system,'' Opt. Lett. \textbf{27}, 258 (2002).

\bibitem{Tai1986} K. Tai, A. Hasegawa, and A. Tomita, ``Observation of modulational instability in optical fibers,'' Phys. Rev. Lett. \textbf{56}, 135 (1986).

\bibitem{Chiao2000} R. Y. Chiao, ``Bogoliubov dispersion relation for a 'photon fluid': Is this a superfluid?,'' Opt. Comm. \textbf{179}, 157 (2000).

\end{thebibliography}
\end{document}